\documentstyle[12pt]{article}

  \evensidemargin=\oddsidemargin
\makeatletter

\let\old@bibitem=\@bibitem
\def\@bibitem#1#2{\@ifundefined{r@#1}{}{\@warning
  {Multiple entries for reference `#1'}}\@ifundefined{b@#1}{\@warning
  {Reference `#1' not cited}}{\global\@namedef{r@#1}{#2}}}
\let\old@lbibitem=\@lbibitem
\def\@lbibitem[#1]#2#3{\@ifundefined{r@#2}{}{\@warning
  {Multiple entries for reference `#2'}}\@ifundefined{b@#2}{\@warning
  {Reference `#2' not cited}}{\global\@namedef{r@#2}{#3}
  \global\@namedef{s@#2}{#1}}}

\def\@putbibrefs#1{\expandafter\@makearef #1*,}
\def\@makearef#1,{\if*#1 \let\next=\relax \else \let\next=\@makearef
  \@ifundefined{s@#1}{\bibitem{#1}}{\bibitem[\@nameuse{s@#1}]{#1}}
  \@ifundefined{r@#1}{Citation label `#1'?}{\@nameuse{r@#1}}
  \fi \next}

\def\endthebibliography{\let\@bibitem=\old@bibitem \let\@lbibitem=\old@lbibitem
  \@putbibrefs\@bibrefslist \endlist}

\def\@citex[#1]#2{\if@filesw\immediate\write\@auxout{\string\citation{#2}}\fi
  \def\@citea{}\@cite{\@for\@citeb:=#2\do
    {\@citea\def\@citea{,\penalty\@m\ }\@ifundefined
       {b@\@citeb}{{\bf ?}\@warning
       {Citation `\@citeb' on page \thepage \space undefined}}%
\hbox{\csname b@\@citeb\endcsname}\@ifundefined
       {d@\@citeb}{\global\@namedef{d@\@citeb}{y}\@addlabel{\@citeb}}{}}}{#1}}

\def\nocite#1{\@ifundefined{d@#1}{\@addlabel{#1}}{\@warning
	{Label `#1' cited.  \string\nocite\space not necessary.}}}

\def\@addlabel#1{\@ifundefined{@bibrefslist}{\xdef\@bibrefslist{#1,}}{\xdef
	\@bibrefslist{\@bibrefslist#1,}}}
\makeatother

\def\sci#1{\ifmmode \times 10^{#1} \else $\times 10^{#1}$ \fi}
\def\etal{{\em et~al.}}

\def\element#1#2{\hbox{${}^#2{\rm #1}$}}
\def\H{\hbox{H}}
\def\D{\hbox{D}}
\def\He#1{\element{He}{#1}}
\def\Li#1{\element{Li}{#1}}
\def\Be#1{\element{Be}{#1}}
\def\B1#1{\element{B}{#1}}

\def\zregion#1#2{\hbox{\rm #1~{\sc #2}}}

\def\hii{\zregion{H}{ii}}

\def\ang{\,{\rm\AA}}

\def\MeV{\,{\rm MeV}}
\def\sec{\,{\rm sec}}
\def\Gyr{\,{\rm Gyr}}

\def\Mpc{\,{\rm Mpc}}

\def\eV{{\,\rm eV}}

\def\cmm2{{\,\rm cm^{-2}}}
\def\cm2{{\,{\rm cm}^2}}
\def\cmm3{{\,{\rm cm}^{-3}}}
\def\gcmm3{{\,{\rm g\,cm^{-3}}}}
\def\kms{\,{\rm km\,s^{-1}}}

\def\la{\mathrel{\mathpalette\fun <}}
\def\ga{\mathrel{\mathpalette\fun >}}
\def\fun#1#2{\lower3.6pt\vbox{\baselineskip0pt\lineskip.9pt
  \ialign{$\mathsurround=0pt#1\hfil##\hfil$\crcr#2\crcr\sim\crcr}}}

\begin{document}

\pagestyle{empty}
\begin{center}
\bigskip
\rightline{FERMILAB--Pub--94/174-A}
\rightline{astro-ph/9407006}
\rightline{submitted to {\it Science}}

\vspace{.35in}
{\Large \bf BIG-BANG NUCLEOSYNTHESIS AND THE\\
\bigskip
BARYON DENSITY OF THE UNIVERSE} \\

\vspace{.2in}
Craig J. Copi,$^{1,2}$ David N. Schramm,$^{1,2,3}$
and Michael S. Turner$^{1,2,3}$\\

\vspace{.2in}
{\it $^1$Department of Physics\\
The University of Chicago, Chicago, IL~~60637-1433}\\

\vspace{.1in}
{\it $^2$NASA/Fermilab Astrophysics Center\\
Fermi National Accelerator Laboratory, Batavia, IL~~60510-0500}\\

\vspace{0.1in}
{\it $^3$Department of Astronomy \& Astrophysics\\
Enrico Fermi Institute, The University of Chicago,
Chicago, IL~~60637-1433}\\

\end{center}
\medskip

\noindent{\bf Big-bang nucleosynthesis is one of the cornerstones of
the standard cosmology.  For almost thirty years its predictions have
been used to test the big-bang model to within a fraction of
a second of the bang.  The concordance that exists between the predicted
and observed abundances of D, $^3$He, $^4$He and $^7$Li provides
important confirmation of the standard cosmology and leads to
the most accurate determination of the baryon density, between
$1.7 \times 10^{-31}\gcmm3$ and $4.1\times 10^{-31}\gcmm3$
(corresponding to between about 1\% and 14\% of critical density).
This measurement of the density of ordinary matter
is crucial to almost every aspect of cosmology and is pivotal to
the establishment of two dark-matter problems:  (i)
most of the baryons are dark, and (ii) if the
total mass density is greater than about 14\% of
the critical density as many determinations now indicate,
the bulk of the dark matter must be ``nonbaryonic,''
comprised of elementary particles left from the earliest
moments.  We critically review the present status of primordial
nucleosynthesis and discuss future prospects.}

\newpage
\pagestyle{plain}
\setcounter{page}{1}

\section{Introduction}

Because of the extremely high temperatures that existed during
the earliest moments it was too hot for nuclei to
exist.  At around $1\sec$ the temperature of the Universe cooled
to $10^{10}\,$K, and a sequence of events
began that led to the synthesis of the light elements
D, $^3$He, $^4$He and
$^7$Li.  The successful predictions of big-bang
nucleosynthesis provide the earliest and most stringent
test of the big-bang model, and together with the expansion of the Universe
and the $2.726\,$K black-body cosmic background radiation
(CBR) are the fundamental observational basis for the standard cosmology.

Big-bang nucleosynthesis began with
the pioneering work of Gamow, Alpher, and Herman who
believed that all the elements in
the periodic table could be produced in the big bang \cite{gamow}; however,
it was soon realized that the lack of stable nuclei of mass 5 and 8 and
Coulomb repulsion between highly charged nuclei
prevent significant nucleosynthesis beyond $^7$Li.  In 1964, shortly
before the discovery of the CBR, Hoyle and Tayler~\cite{fredht}
argued that the big bang must produce a large $^4$He abundance (about 25\% by
mass) and this could provide the explanation for the high \He4 abundance
observed in many primitive objects.  At about the same time,
Zel'dovich realized that a lower temperature for
the Universe today implied a greater mass fraction of \He4 produced
in the big-bang, and concluded that the big-bang model was in trouble.
While his reasoning was correct, through a comedy of
misunderstandings he mistakenly believed
that the primeval mass fraction of $^4$He was
at most 10\% and that the temperature of the Universe
was less than about $1\,$K \cite{zel}.

After the discovery of the CBR by Penzias and Wilson
in 1965, detailed calculations of big-bang nucleosynthesis
were carried out by Peebles~\cite{pjep-bbn} and by
Wagoner, Fowler and Hoyle~\cite{wfh}.  It soon became clear that, as
Hoyle and Tayler had speculated, the origin of the large
primeval fraction of $^4$He was indeed the big-bang, and
further, that other light elements were also produced.
However, the prevailing wisdom was
that \D\ and \Li7 were produced primarily during the T~Tauri phase of stellar
evolution and so were of no cosmological significance \cite{fgh}.
The amount of \He4 produced in the big bang is very insensitive to
the cosmic baryon---that is, ordinary matter---density (see Fig.~1),
and thus it was not possible to reach any
conclusions regarding the mean density of ordinary matter.

Since the other light elements are produced in much smaller quantities,
ranging from $10^{-5}$ or so for D and $^3$He to $10^{-10}$ for
$^7$Li (see Fig.~1), establishing their big-bang origin was a much
more difficult task.  Further, it is complicated by the fact
that the material we see today has been subjected to more
than $10\Gyr$ of astrophysical processing, the details of which are
still not understood completely.  However, over the past 25 years the
big-bang origin of D, $^3$He, and $^7$Li has been established,
not only further testing the model, but also enabling an
accurate determination of the average density of baryons in the Universe.

First, it was shown that there is no plausible astrophysical site for the
production of deuterium~\cite{rafs,els}; due to its fragility, post big-bang
processes only destroy it.  Thus, the presently observed deuterium
abundance serves as a {\it lower limit\/} to the big-bang production.
This argument, together with the strong dependence
of big-bang deuterium production on the baryon density, led to the
realization that D is an excellent ``baryometer''~\cite{rafs,ggst},
and early measurements of the deuterium abundance~\cite{gr,ry},
a few parts in $10^5$ relative to hydrogen, established
that baryons could not contribute
more than about $20\%$ of closure density.  This important conclusion
still holds today.

The chemical evolution of $^3$He is more complicated.
Helium-3 is produced in stars as they burn their
primeval D before reaching the main sequence,
and later by the nuclear reactions
that cook hydrogen into helium.  Some massive stars
destroy (or astrate) $^3$He.  It wasn't until the
late 1970's that a suitable argument for $^3$He was
found:   The present sum of D + $^3$He bounds
their combined big-bang production \cite{ytsso}.

Lithium was the last to come into the fold.  Stellar processes
both destroy and produce $^7$Li; moreover, the abundance of
$^7$Li varies greatly, from $^7$Li/H$\,\simeq 10^{-9}$ in the
interstellar medium (ISM) to less than $^7$Li/H$\,\simeq 10^{-12}$
in some stars.  In 1982, Spite and Spite circumvented these
difficulties by measuring the \Li7 abundance
in the oldest stars in our galaxy, metal-poor, pop II halo stars.
They found $^7$Li/H$\,\simeq 10^{-10}$ \cite{ss}, which
is consistent with big-bang production.
Their results established the case for the primeval \Li7
abundance, which since has been strengthened
by the work of others~\cite{rmb-hp,thorburn}.

For the last decade much effort has been devoted to the
critical comparison of the theoretical predictions and
inferred primordial abundances of the light elements.
The predictions depend upon the ratio of baryons to
photons ($\equiv \eta$).  As we shall discuss in more detail, if
$\eta$ is between about $2.5\times 10^{-10}$ and $6\times 10^{-10}$
there is concordance between the predicted and measured abundances of
all four light elements (see Fig.~1).  This leads to the
best determination of the baryon density,
\begin{equation}
\rho_B = \eta n_\gamma m_N =
1.7\times 10^{-31} \gcmm3 \hbox{--} 4.1 \times 10^{-31}\gcmm3 ,
\end{equation}
where the number density of photons, $n_\gamma = 411 \cmm3$,
is known very precisely because the CBR temperature is
so well determined~\cite{firas}, $T_0 = 2.726\,{\rm K} \pm 0.005\,{\rm K}$.
On the other hand, because the critical density,
\begin{equation}
\rho_{\rm crit} = 3H_0^2/8\pi G = 1.88 h^2 \times 10^{-29}\gcmm3 ,
\end{equation}
depends upon the Hubble constant, which is still only known to
within a factor of two,
the fraction of critical density contributed by baryons is less well known:
\begin{equation}
\Omega_B = 0.009h^{-2} \hbox{--} 0.02h^{-2} ,
\end{equation}
where $h\equiv H_0/100\kms\Mpc^{-1}$.
For a generous range for the Hubble constant, $h=0.4 - 1$,
baryons contribute between 1\% and 14\% of closure density.

This fact has two profound implications.  First, since ``optically''
luminous matter (stars and associated material)
contributes much less than 1\% of the critical density,
$\Omega_{\rm LUM} \approx 0.003h^{-1}$ \cite{omega-vis},
most baryons must be dark, e.g., in the form of hot,
diffuse gas, or ``dark stars'' that have either
exhausted their nuclear
fuels (black holes, neutron stars or white dwarfs) or
were not massive enough (less than about $0.08M_\odot$)
to ignite them.   In clusters of galaxies most of the
baryonic matter seems to be in the form of hot, x-ray
emitting gas.  Further, there is now indirect
evidence for the existence of dark stars,
known as MACHOs for Massive Astrophysical Compact
Halo Objects, through their gravitational microlensing of
distant stars \cite{macho}.

Second, there is strong---though
not yet conclusive---evidence that the average mass density of
the Universe is significantly greater than 14\% of the critical
density \cite{omega0}; if this is indeed the case, most of the mass density
of the Universe must be ``nonbaryonic,'' with the most promising
possibility being elementary particles left over from the
earliest moments of the Universe \cite{particledm}.
Large-scale experiments are underway in laboratories
all over the world to directly detect the nonbaryonic
dark matter associated with the halo of our own galaxy \cite{dmdetect}.

Big-bang nucleosynthesis
plays the central role in defining both dark-matter problems
which are central to cosmology today.  For example,
the detection of temperature variations in the CBR by the
COBE satellite was a dramatic confirmation of the general
picture that structure evolved from small density inhomogeneities
amplified by gravity.  One of the great challenges in cosmology
is to formulate a coherent and detailed
picture of the formation of structure (i.e., galaxies, clusters
of galaxies, superclusters, voids, and so on) in the Universe;
the nature of the dark matter is crucial to doing so.

Primordial nucleosynthesis also allows us to ``study'' conditions
in the early Universe, and thereby, to probe fundamental
physics in regimes that are beyond the reach of terrestrial
laboratories.   For example, more than ten years ago
the overproduction of $^4$He was used
to rule out the existence of more than three light (mass less
than about $1\MeV$) neutrino species and constrain the existence of
other light particle species \cite{ytsso,nulimit,ossw,wssok}.

The remainder of this article is given to a careful assessment
of the predictions and observations, paying special attention
to the conclusions that can be sensibly draw about the baryon density.
We begin with the easier part, a discussion
of the theoretical predictions, where the few uncertainties are primarily
statistical in nature and easy to quantify.  We then move on
to the more difficult side, observations.  Here the situation is
just the reverse:  The uncertainties are dominated by possible systematic
errors and interpretational issues, which cannot be characterized
by standard Gaussian error flags; care and judgment must be
exercised to reach reliable conclusions.

\section{Theoretical expectations}

Things have come a long way from the early semi-quantitative
estimates of Gamow and his collaborators \cite{gamow}.  The first serious
attempts to calculate big-bang $^4$He production were those of
Alpher, Follin and Herman~\cite{afh} and Hoyle and Tayler~\cite{fredht}.
After the discovery of the CBR, Peebles~\cite{pjep-bbn}, and then
Wagoner, Fowler, and Hoyle~\cite{wfh}, carried out detailed
numerical calculations incorporating a reaction network of more
than ten nuclides (see Fig.~2).
The code developed by Wagoner et al.~\cite{wfh,code-orig} (using
numerical techniques introduced by Truran~\cite{truran} to
study nucleosynthesis in exploding stars) has since been modified
to incorporate small, but significant corrections
(finite-temperature effects, finite nucleon-mass effects,
the slight coupling of neutrinos to the electromagnetic plasma,
and so on), to update nuclear-reaction rates, and to improve the
numerical-integration scheme.  Its present-day incarnation
has become the standard code (though other codes have been
developed independently), and is described in detail in
Refs.~\cite{code,bbnrev}.

The assumptions underlying the standard scenario of big-bang
nucleosynthesis are few:  (i)
Big-bang (i.e., Friedmann-Robertson-Walker) cosmological model;
(2) three massless (or very light) neutrino species; (3) small or
vanishing neutrino chemical potentials; (4) no additional
light particle species present in
thermal abundance; (5) spatially uniform baryon-to-photon ratio.
In addition, there are ``nuclear input parameters:''
neutron mean lifetime, which determines the matrix element for
all the reactions that interconvert neutrons and protons, and
cross sections for nuclear reactions.

As recently as ten years ago the uncertainty
in the mean neutron lifetime was significant; at present it
is known very precisely:  $\tau_n = 889\sec \pm 2\sec$ \cite{neutron}.
The other cross sections that are required have been measured in the
laboratory at energies appropriate for primordial nucleosynthesis
(this is in contrast to stellar nucleosynthesis where
lab-measured cross sections must be extrapolated to much
lower energies).  With the exception of \Li7,
the uncertainties in cross sections
do not result in significant uncertainties in the light-element yields.

Two recent Monte-Carlo studies have quantified
the uncertainties in the predicted
abundances~\cite{bbnrev,kk}.
Kernan and Krauss~\cite{kk} ran a suite of 1000 models with input parameters
chosen from the probability distributions for the
various cross sections and neutron mean lifetime.
For a baryon-to-photon ratio of $3\times 10^{-10}$
the ``two-sigma'' range of the abundances found
was:  $Y_P = 0.239\hbox{--}0.241$;
$\D/\H=6.7\times 10^{-5}\hbox{--}9.0\times 10^{-5}$; $\He3/\H=1.4\times
10^{-5}\hbox{--}1.9\times 10^{-5}$; and $\Li7/\H=0.81\times 10^{-10}\hbox{--}
1.7\times 10^{-10}$ (i.e., 950 of the models had
abundances within the stated intervals).  Here, $Y_P$ is
the mass fraction of $^4$He produced; the other abundances
are specified by number relative to hydrogen.
Only for $^7$Li is the uncertainty
significant when compared to the uncertainty in the
observed abundance.  It arises due to three cross sections that
are still poorly known:  $\He3+ \He4 \rightarrow \gamma +\Be7$,
${}^3{\rm H} + \He4 \rightarrow \gamma + \Li7$, and $\Li7 + p
\rightarrow \He4 + \He4$.  (In fact,
there could be systematic errors in one or more of these
cross sections, as the results of different experiments are
not consistent with their quoted error flags \cite{bbnrev,kk}.)

As an aside, the $^4$He yield is known most accurately, apparently
to a precision of better than 1\%.  For this
reason even tiny corrections have become important, and it
is difficult to judge whether or not every significant effect
has been taken into account.  The most recent correction
may serve as a guide:  finite nucleon-mass effects led to an increase,
$\Delta Y_P \simeq +0.001$ \cite{finitenucmass}.
Further, an informal poll of the
various nucleosynthesis codes known to us gave results that, with the
exception of $^4$He, fell within the above mentioned Monte-Carlo range;
for $^4$He, the values reported ranged from $Y_P = 0.237$ to $Y_P = 0.241$.

Modifications of the standard scenario have also been investigated, including
almost every imaginable possibility \cite{mathewsmalaney}:  additional light
particle species; unstable, massive tau neutrino;
decaying particles; variations in the fundamental
constants; large neutrino chemical potentials; primeval magnetic
fields; and spatial variations in the baryon-to-photon ratio.
In most instances the ``nonstandard physics''
was introduced for the purpose of obtaining
a bound or limit based upon primordial nucleosynthesis, e.g.,
the previous mentioned limit to the number of light neutrino species.
In a few cases, however, it was motivated by other considerations.

For example, Witten suggested that if the transition from
quark/gluon plasma, which existed prior to about $10^{-5}\sec$,
to hadron matter involved a strongly first-order phase transition
the resulting distribution of baryons could be quite
inhomogeneous \cite{witten}, thereby possibly significantly changing
the outcome of primordial nucleosynthesis.  For a short time, it
appeared that such inhomogeneity could lead to a relaxation
of the bound to $\Omega_B$,
even permitting closure density in baryons~\cite{ahs-bk}.  It is now clear
that any significant level of inhomogeneity upsets the agreement
of the predictions with the observations~\cite{kmos,st,tsommf};
moreover, there is now little motivation from particle physics
for a strongly first-order quark/hadron phase transition.
At present, the only modification involving the known
particles that leads to significant changes is the
possibility that the tau neutrino has a mass of the order
of $1\MeV \hbox{--} 30\MeV$ \cite{massivetau}.  The present laboratory limit
to its mass is just above $30\MeV$ and should be
improved enough to clarify this issue soon.

To summarize the theoretical situation; the only compelling scenario
for primordial nucleosynthesis is the standard one.  Within the standard
picture the predictions for the light-element abundances
have uncertainties that, with the exception of $^7$Li, are
not significant when compared to the accuracy with which the
primeval abundances are known.  At present,
the comparison between theory and observation
turns primarily on the observations, and there, the uncertainties
are more difficult to quantify.

\section{Confrontation Between Theory and Observation}

The predictions of the standard scenario, including ``two-sigma''
uncertainties based upon our Monte-Carlo calculations
are shown in Fig.~1.\footnote{Our Monte-Carlo calculations are similar
to those in Refs.~\cite{bbnrev,kk} with one exception; we have
treated the cross sections for \Li7 production differently.
For the cross sections where data from two experiments are
inconsistent we have used both distributions, alternating
between the two.}
We now discuss the {\it inferred}\/ primordial abundances, with emphasis
on {\it inferred,} since one must deduce the primordial
abundances of D, $^3$He, $^4$He, $^7$Li from material that
has undergone some 10 Gyr of chemical evolution.

\subsection{Deuterium and Helium-3}

Since deuterium is the most weakly bound, stable nucleus it is easy to destroy
and difficult to produce.  As discussed earlier,
the deuterium abundance observed today
provides a lower limit to the big-bang production.  The
Apollo Solar Wind Composition experiment, which captured solar-wind particles
in foils exposed on the moon, and the subsequent analysis by Geiss and
Reeves~\cite{gr} provided the first accurate assessment of the pre-solar \D\
and \He3 abundances.  Based on these experiments and studies
of primitive meteorites,
Geiss deduces a pre-solar (i.e., at the time of the formation
of the solar system) deuterium abundance~\cite{geiss}
\begin{equation}
 \left( \frac{\D}{\rm H} \right)_\odot = 2.6 \pm 1.0 \sci{-5}.
\label{eqn:y2ps}
\end{equation}
This value is consistent with measurements of the deuterium abundance
in the local ISM (i.e., within a few hundred pc)
made two decades ago by the Copernicus
satellite~\cite{ry}, and more recently by the Hubble Space
Telescope~\cite{linsky},
\begin{equation}
 \left( \frac{\D}{\rm H} \right)_{\rm HST} = 1.65^{+0.07}_{-0.18} \sci{-5}.
\end{equation}
That the ISM value is slightly lower than the pre-solar
abundance is consistent with slow depletion of deuterium with time since
the material in the ISM is about 5~Gyr younger than the material from which
our solar system was assembled.   A sensible lower bound to the primordial
deuterium abundance,
\begin{equation}
 \left( \frac{\D}{\rm H} \right)_P \ge 1.6\sci{-5}, \label{eqn:y2p}
\end{equation}
based on these measurements
leads to an upper limit to $\eta$ of $9\times 10^{-10}$.
Because of the rapid variation of the amount of deuterium produced
with $\eta$, this upper limit is rather insensitive to the exact
lower bound adopted for D/H.  Further, this argument is very
robust because it involves
minimal assumptions about galactic chemical evolution,
simply that \D\ is only destroyed by stellar processing \cite{els}.

It would be nice to exploit the rapid variation of deuterium production
to obtain a lower bound to $\eta$, based upon the overproduction
of deuterium.  This cannot be done directly
because deuterium is so easily destroyed.
However, an equally useful bound can be derived based upon the
sum of D + $^3$He production.  Primordial deuterium either resides in
the ISM or has been burnt to \He3 (via
D + $p\rightarrow \gamma$ + $^3$He).  A significant fraction
of \He3 survives stellar
processing (in fact, low mass stars are net producers of
$^3$He), and thus an upper bound to the primordial
\D\ + \He3 abundance can be inferred from present-day measurements with
few assumptions.  This line of reasoning was introduced
by Yang et al.~\cite{ytsso} who derived the bound,
\begin{equation}
 \left( \frac{\He3 + \D}{\rm H} \right)_P \le  \left( \frac{\He3 + \D}{\rm H}
\right)_\odot + \left( g_3^{-1} - 1 \right) \left( \frac{\He3}{\rm H}
\right)_\odot,
\end{equation}
where the \He3 survival fraction $g_3$ was argued to be
greater than 25\%~\cite{dss}.  (It should be noted that
even massive stars, which tend to burn \He3, eject some \He3
in their winds.)  The bound was improved by taking account of material
that has been processed by more than
one generation of stars~\cite{ossw,wssok}.  Both
methods lead to similar upper limits to the primordial $\D+\He3$ abundance,
\begin{equation}
 \left( \frac{\D+\He3}{\rm H} \right)_P \le 1.1 \sci{-4},\label{eqn:y23p}
\end{equation}
which in turn gives the bound $\eta\ge 2.5\times 10^{-10}$.  Like
the upper limit to $\eta$ based upon deuterium, this lower limit
is insensitive to the precise bound to the primeval abundance of D + \He3
because of the steep rise of D + \He3 production with
decreasing $\eta$.  Together, D and \He3 define a concordance
interval, $2.5\times 10^{-10}\le \eta \le 9\times 10^{-10}$.

The theoretical belief that low-mass stars actually increase
the D + $^3$He abundance by producing \He3 is supported by
the observations of Wilson, Rood and Bania~\cite{wrb}.
By using the analogue of the 21\,cm hydrogen hyperfine
transition for $\He3^+$ they found $\He3 /
{\rm H}\sim 10^{-3}$ in planetary nebulae.  This much additional
$^3$He production agrees with the value predicted by
stellar models of Iben and Truran~\cite{it}.  However,
measurements of the \He3 abundance by the same method in hot, ionized
gas clouds, so called \hii\ regions, vary greatly, from
$\He3/{\rm H} \sim 1\times 10^{-5}$ to $\He3/{\rm H}\sim
8\times 10^{-5}$ \cite{he3},
which suggests that \He3 is destroyed
by varying degrees \cite{he3-dest}.  While \hii\ regions are the
only place outside the solar system where the \He3 abundance can be measured,
they are samples of the cosmos dominated by the effects of
massive, young stars, which are
the most efficient destroyers of \He3, and thus, they do not represent
``typical samples'' of the cosmos
(so far as the chemical evolution of \He3 is concerned).
In any case, we believe that a \He3 survival fraction of
$25\%$ or more remains a reasonably conservative estimate as
applied to the solar system \He3 abundance.

\subsection{Helium-4}

In two important regards the primordial \He4 abundance is the easiest to
measure:  it is large, around 25\% by mass fraction, and the chemical
evolution of \He4 is straightforward---stars are net producers of \He4.
On the other hand, the predicted abundance is most accurately
known and varies only logarithmically with $\eta$.  Thus,
measuring the \He4 abundance to sufficient accuracy to sharply
test the big-bang prediction is just as
challenging as determining the other light-element abundances.

Needless to say, since \He4 is ubiquitous, its abundance can be
measured in many different ways, all of which give values consistent
with a primeval mass fraction of around 25\%.  The most
accurate determinations of the primeval \He4 abundance rely on measurements of
its recombination radiation in low-metallicity
\hii\ regions (see Ref.~\cite{dk} for a detailed discussion of the
experimental method).  Since stars produce both helium and other
heavier elements, contamination due to stellar production
should be minimized in metal-poor samples of the Universe.
A number of groups have obtained high-quality data for very
metal-poor, extragalactic \hii\ regions, which has allowed
determination of helium abundances to very good statistical
accuracy~\cite{he4-meas}.  Moreover, several independent
and detailed analyses of these data sets
have been carried out~\cite{wssok,pagel,ost}.
The quality of the data and the accuracy of the abundance
determinations desired are now such that possible
systematic errors dominate the error budget,
and they are the focus of our attention.

\subsubsection{Systematic Effects}

The first step in the path to the primordial helium abundance is
measuring line strengths of the recombination radiation for
hydrogen and helium.  Line strengths
are then translated into a helium
mass fraction by means of theoretical emissivities for
both helium and hydrogen and modeling of the \hii\ region.  In modeling an
\hii\ region spherical symmetry and uniform temperature
are assumed, neither of which is an excellent
assumption given that a typical \hii\ region is heated
by a few massive, young stars near its center.  Since the
ionization potentials for hydrogen and helium are
different, corrections must be made for both neutral and doubly ionized helium.
Collisional excitation can be significant, but is not easy to
accurately estimate.  Stellar absorption by the stars heating the \hii\
region can affect the excitation of the hydrogen and
helium in the \hii\ region.  Absorption due to intervening dust
can also affect the abundance determinations.
A summary of our estimate of the systematics, based largely on the discussion
of Skillman and Kennicutt~\cite{sk} and Skillman et al.
\cite{skill94}, is given
in Table~\ref{tab:sys-errors}.  A detailed numerical assessment of
some of these effects has recently been carried out
by Sasselov and Goldwirth~\cite{y-system}, who suggest that
the systematic errors may even be slightly larger.

\begin{table} \center
\begin{tabular}{p{4in}r} \hline\hline
  Type of correction & Estimate \\ \hline
  Line ratios (including dust absorption) & $\pm 2\%$ \\
  Emissivities & $\pm 2\%$ \\
  Collisional excitation and stellar absorption & $\pm 1\%$ \\
  Neutral helium & $+2\%$ \\ \hline
  Total & $+7\%$,$-5\%$ \\ \hline\hline
 \end{tabular}
 \caption{Estimate of systematic errors.}
 \label{tab:sys-errors}
\end{table}

\subsubsection{Primordial helium-4 abundance}

Even in the most metal-poor \hii\ regions some of the
\He4 is produced by stars.  Since stars also produce
the elements beyond \He4 (collectively referred to as
metals), there should be a direct relationship between
metallicity and stellar-produced \He4.  Peimbert and
Torres-Peimbert~\cite{ptp} pioneered the extrapolation
of the helium abundance vs. heavy-element abundance
to zero metallicity to infer the primordial \He4 abundance.
Oxygen, nitrogen, and carbon have
all been used as indicators of stellar nucleosynthesis and hence
the amount of stellar produced \He4.  Each has its advantages and
disadvantages~\cite{sgs}, though the quantitative results are very similar.
Recently, Olive and Steigman~\cite{ost} have performed a detailed statistical
analysis of very metal-poor \hii\ regions, and
derive a primordial \He4 abundance (see Fig.~3)
\begin{equation} Y_P = 0.232 \pm 0.003^{+0.016}_{-0.012} \end{equation}
where their statistical error is quoted first and the systematic error
based upon Table~\ref{tab:sys-errors} appears second.  (For reference,
their quoted systematic error is $\pm 0.005$.)

To summarize, there is undisputed evidence for a large
primeval \He4 abundance whose only plausible explanation
is the big bang.  Following Olive and Steigman \cite{ost}
we take as a reasonable estimate for the primeval mass fraction,
$Y_P = 0.221 - 0.243$,
which allows for a two-sigma statistical uncertainty and
their one-sigma systematic uncertainty.  Within the
two-sigma theoretical uncertainty, such a primeval
mass fraction of \He4 is consistent
with the big-bang prediction provided $0.8 \times 10^{-10} \le \eta
\le 4 \times 10^{-10}$.  At present, errors
are dominated by possible systematic effects; allowing for
our higher estimate of systematic error, a primeval
\He4 mass fraction as low as 0.214, or as high as 0.254, cannot
be ruled out with certainty.  This extreme range for
the primeval \He4 abundance is consistent with a much larger
interval, $6\times 10^{-11} \le \eta \le 1.5\times 10^{-9}$,
illustrating the exponential sensitivity of $\eta$ to $Y_P$.

\subsection{Lithium}

The study of extremely metal-poor, pop II halo stars has
provided the bulk of our knowledge of the light elements beyond
\He4.  It began with the work of Spite and Spite~\cite{ss}, who
measured the \Li7 abundance as a function of metallicity
(iron abundance) and surface temperature.
Much to their surprise they found a ``plateau'' in the \Li7
abundance and established what has become
a very strong case for the determination of the primeval \Li7 abundance.

The Spite plateau refers to the fact that the \Li7 abundance as a function
of surface temperature is flat for surface temperatures greater
than about 5600K (see Fig.~4), and further that the \Li7
abundance for stars with temperatures greater than 5600K as a
function of iron abundance is flat for very low iron abundance (Fig.~5).
The first plateau gives strong evidence that the stars with the
highest surface temperatures are not destroying their \Li7
by convective burning since the depth of the convective
zone depends upon the surface temperature (for temperatures
lower than 5600K the measured \Li7 does vary with surface
temperature indicating convective burning).  The second plateau
indicates that any \Li7 due to stellar production
must be insignificant for the most metal-poor stars since
the \Li7 abundance is independent of the metal abundance.

The actual value of the \Li7 abundance on ``the Spite plateau'' is subject to
several important systematic effects.  In particular, model atmospheres
used by different authors assume effective surface temperatures,
differing by as much as $200\,{\rm K}$\@.  Other differences
in the model atmospheres, including assumptions made about opacities,
also affect the inferred \Li7 abundances in a systematic way.
These systematic effects explain the main difference between the
Spite and Spite abundance, ${\rm \Li7/H} = 1.1\sci{-10}$, and the value
derived recently by Thorburn~\cite{thorburn} from a sample of 90
pop II stars, ${\rm \Li7/H} = 1.7\sci{-10}$\@ (see Figs.~4, 5).
Further, Thorburn's data seems to indicate a slight
systematic variation of the \Li7 abundance
with surface temperature, possibly indicating some depletion from a higher
primordial value by processes that transport \Li7 inward to
regions of high enough temperature that it can be burned;
e.g., meridonal mixing~\cite{mixing}.  However, the amount of depletion
is constrained by the relatively narrow spread in \Li7 abundance
for a wide range of surface temperatures and metallicities.
Microscopic diffusion is ruled out by this fact, though
stellar models that incorporate rotation, which suppresses
microscopic diffusion, can be made consistent with the observations~\cite{cd}.

The case against significant depletion (and hence for a primeval
abundance) was further strengthened by the observation of
\Li6 in a pop II star by Smith, Lambert, and Nissen~\cite{sln};
Hobbs and Thorburn~\cite{ht} have detected \Li6 in this and
another pop II star.  Big-bang production of \Li6 is negligible and so the \Li6
seen was
presumably produced by cosmic-ray processes, along with beryllium
and boron (as discussed below).  Since \Li6 is much more fragile
than \Li7 and yet still survived with an abundance relative
to Be and B expected for cosmic-ray production, depletion of \Li7
cannot have been very significant~\cite{os,sfosw}.
These \Li6 measurements allow for a largely model-independent
limit to the amount of \Li7 depletion, less than about a factor of two.

In summary, based on metal-poor, pop II halo stars we infer a
primordial \Li7 abundance of
\begin{equation}
 \left( \frac{\Li7}{\H} \right)_P = 1.4 \pm 0.3^{+1.8}_{-0.4} \sci{-10}
\end{equation}
where the central value is the average of the Spite and Spite
and Thorburn determinations, the statistical error is listed first,
and the systematic error second.  The systematic-error
estimate consists of $\pm 0.4$ due to differences in
model atmospheres and $+1.4$ to account for possible depletion.
In fixing a range for the primordial \Li7 abundance it is
the systematic error that is most important; accordingly,
we use the sum of statistical plus systematic error to derive
our estimate for the \Li7 abundance, $0.7\times 10^{-10}
\le \Li7/{\rm H} \le 3.5\times 10^{-10}$.  Allowing for the
two-sigma theoretical uncertainty, the concordance interval
is $1 \times 10^{-10} \le \eta \le 6\times 10^{-10}$.  (We note that the
95\% confidence range for the \Li7 abundance advocated by
Thorburn \cite{thorburn} differs only slightly from ours.)

\subsection{Beryllium and Boron}
While the inhomogeneous variant of big-bang nucleosynthesis
motivated by a first-order quark/hadron phase transition
cannot alter the basic conclusions, an important question remains, namely, is
there an observable signature that can differentiate between the
inhomogeneous and the homogeneous models, thereby probing the
quark/hadron transition? Several authors~\cite{ahs-bk} argued that
the regions with high neutron-to-proton ratio that exist
in inhomogeneous models could lead to ``leakage'' beyond
mass 5 and mass 8 and traces of \Be9, $^{10}$B,
$^{11}$B, and possibly even r-process elements (neutron-rich
isotopes) could be produced.
However, detailed studies by Sato and Terasawa~\cite{st} and
Thomas et al.~\cite{tsommf} have shown that such leakage is negligible
when the D, \He3, \He4, and \Li7 abundances are consistent
with their observed values, with Be/H, B/H $\sim 10^{-18}$

An observational program similar to that of Spite and Spite~\cite{ss}
was begun for beryllium and boron.  Recently, both
beryllium~\cite{be-meas} and boron \cite{dll} have been
detected in metal-poor, pop II halo stars.  The
observations indicate that beryllium and boron abundances
scale with metallicity, strongly suggesting that their
production was not the big bang \cite{beb}.
The processes that produce the beryllium and boron (and \Li6)
seen in younger pop I stars (like our sun)
are thought to be cosmic-ray reactions~\cite{rfh}.
For Be and B, such reactions involve the breakup of heavy nuclei such as
C, N, O, Ne, Mg, Si, S, Ca, and Fe by protons and alpha particles
(for lithium in pop II stars, alpha plus alpha fusion reactions are
dominant~\cite{fusion}).

\subsection{Toward truly primordial abundances}

As the reader by now should appreciate, the task of disentangling
10\,Gyr of galactic chemical evolution is not an easy one.  What are
the prospects for determining the light-element abundances
in very primitive samples of the Universe (that is, in
objects seen at very high red shift)?

Hydrogen clouds at high redshift ``backlit'' by quasars
offer the possibility of measuring the deuterium
abundance in very old, very distance, and very primitive samples of
the cosmos~\cite{webb}.  These clouds, known as quasar absorption
line systems, are ``seen'' by the absorption features they
produce in the quasar spectrum; many are observed to be
very metal-poor, indicative of primeval material.  There have been many
searchs for the deuterium analog of the $1216\ang$ Lyman-$\alpha$
absorption feature, which is shifted very slightly to the blue,
by about $0.33\ang$.  Recently, Songaila et al. and Carswell et
al.~\cite{qso-d}
announced a possible detection of deuterium in a redshift
$z=3.32$ absorption line system (in the quasar Q0014+813);
if it is deuterium, it corresponds to an abundance
\begin{equation}
 \left( \frac{\D}{\rm H} \right)_{\rm abs} = (1.9 \hbox{--} 2.5) \sci{-4}.
\end{equation}
Both groups are quick to point out that a single measurement
does not constitute a definite detection of deuterium as there
is a significant probability ($\sim 15\%$) that the feature
seen arises from Lyman-$\alpha$ absorption due to
another, smaller hydrogen cloud at slightly lower
redshift.  At the very least though, their detection provides
an important upper limit to the deuterium
abundance in this very primitive sample of the cosmos.

Interpreting the detection as an upper bound to the primordial
deuterium abundance leads to the constraint $\eta \ge 1.6\times
10^{-10}$, only slightly less stringent than
the previous bound based upon the production of D + $^3$He.
If, on the other hand, it is interpreted as a measurement
of the primordial deuterium abundance,
then $\left( \D/{\rm H} \right)_P \sim 2\times 10^{-4} \gg
\left[ ( \D+\He3)/
{\rm H} \right]_\odot \sim 4\times 10^{-5}$, which leads to a
problem in understanding the observed pre-solar D+\He3 abundance.
Since it is almost certain that \D\ is
destroyed by burning to \He3 one would expect a
much higher D + \He3 abundance than has been observed.
This could indicate a problem with models of the chemical evolution of
$^3$He, or the interpretation as a deuterium detection~\cite{steigman}.

Carswell et al. have studied another quasar at a similar
redshift (Q0420-388) and have detected deuterium in an absorption
line system at the level of
about $2\times 10^{-5}$, with a three-sigma upper limit of
$6\times 10^{-5}$ \cite{newdeuterium}.
This new observation seems to imply that the
previous detection was really due to another small hydrogen cloud.
Further, if correct, it fits nicely with the local
measurements of D and \He3, and suggests that the astration of
\D\ is not great.

The merits of using quasar absorption line systems to obtain
the primordial deuterium abundance are clear.  The means necessary
for such observations are
now at hand:  large-aperture telescopes with very high-resolution
spectrometers, such as the 10\,meter Keck telescope used by
Songalia et al.  With some luck, it should just be a matter of
time before the deuterium
Lyman-$\alpha$ feature is measured in several such systems.  Once
it is, and if the abundances are similar, both the reality
of the feature and its interpretation as reflecting the primordial
D abundance will have been established.

With regard to \He3, one might hope to eventually use the
$\He3^+$ hyperfine line to determine the \He3 abundance
in extragalactic \hii\ regions that are very metal poor.
However present technology is only marginally sufficient
to observe galactic \He3 so it will take time
before extragalactic detections are possible.

The \He4 abundance has been measured through its absorption
lines in a quasar at redshift $z=2.72$ (HS1700+6414) \cite{qsohe4},
and, very recently, observations made with the refurbished
Hubble Space Telescope have revealed the presence of singly
ionized \He4 in the intergalactic medium \cite{jakobsen}.
While both measurements provide important confirmation of
a large, primeval \He4 abundance in very primitive samples
of the cosmos, they lack the precision
necessary to sharply test big-bang nucleosynthesis.  In
that regard, metal-poor extragalactic \hii\ regions provide
the most accurate determinations.

Finally, owing to its small abundance, it seems very unlikely
that the \Li7 abundance can be measured in high-redshift objects,
or even in extragalactic stars.  On the other hand, the data at
hand present a good case for having determined the \Li7
abundance in the very oldest stars in our galaxy.

\section{Implications and Future Directions}

The agreement between the predictions of primordial
nucleosynthesis and the inferred primordial abundances is
impressive---all the more so when viewed
in light of the sharpening of the theoretical predictions and
the improvement in the observational data that the past decade has
witnessed.  Without a doubt, primordial nucleosynthesis provides the
most significant test of the standard cosmology---which it
passes with flying colors---and leads
to the best determination of the density of ordinary matter.

Where do we stand?  The data are not yet good
enough to single out a value for the baryon-to-photon
ratio.  However, they are good enough to delineate a very narrow
``concordance interval'' where the predicted abundances of
all four light elements are consistent with their measured values.

The lower limit to the concordance interval hinges primarily
upon the D + $^3$He abundance.  Based upon our understanding
of the difficulty of efficiently destroying $^3$He,
$\eta = 2.5\times 10^{-10}$ stands as a reliable lower bound.
This lower bound is buttressed by both $^7$Li---for
$\eta \le 1 \times 10^{-10}$ the predicted $^7$Li abundance rises above $3.5
\times 10^{-10}$---and by the Songaila et al. upper limit to the
primitive deuterium abundance---for $\eta \le 1.6\times
10^{-10}$ D/H exceeds $2.5\times 10^{-4}$.\footnote{Had
we used the Carswell et al. \cite{newdeuterium} upper limit,
D/H$\,\le 6\times 10^{-5}$, which has yet to be published,
the lower bound to $\eta$ would be very similar to that
based upon D + \He3.}

The upper limit to the concordance interval derives from $^4$He,
$^7$Li and D, with the stringency of the limits in that order,
but the reliability in the reverse order.  Assuming that the
primordial mass fraction of $^4$He is no larger than 0.243,
based upon the work of Olive and Steigman \cite{ost}, then
$\eta$ must be less than $4\times 10^{-10}$.  On the other
hand, if owing to systematic error $Y_P$ is as large as 0.254,
then $\eta$ could be as large as $1.5\times 10^{-9}$.
This illustrates the point mentioned earlier, the upper limit to $\eta$
depends exponentially upon the upper limit to $Y_P$,
\begin{equation}\label{eq:exp}
\eta_{\rm max} \simeq 4\times 10^{-10} \, \exp
[100 (Y_P^{\rm max}-0.243)] .
\end{equation}
While \He4 is arguably the most striking confirmation of
big-bang nucleosynthesis, the logarithmic dependence of
the \He4 mass fraction on $\eta$ makes it a very poor baryometer.

The uncertainty in our reasonable upper bound to \Li7, \Li7/H$\,\le
3.5\times 10^{-10}$, is primarily systematic error
associated with possible \Li7 depletion in metal-poor, pop II stars.
Our upper bound to \Li7 implies $\eta \le 6\times 10^{-10}$.
On the other hand, since the strongest argument against very
significant depletion of $^7$Li in metal-poor, pop II stars
is the observation of \Li6, which has only been seen in
two stars, very significant depletion of \Li7 cannot be ruled
out with certainty.  Taking as an extreme upper limit,
$^7$Li/H$\,\sim 6\times 10^{-10}$, corresponding to a factor
of four depletion, $\eta$ could be as large as $9 \times 10^{-10}$.

Finally, turning to deuterium; because it is so easily destroyed
and lacks a plausible contemporary astrophysical site for its origin,
its primordial abundance must be larger than what is seen
today:  D/H\,$\ge 1.6\times 10^{-5}$.  This implies an upper bound
to $\eta$ of $9 \times 10^{-10}$.  It seems very difficult to get around
this simple argument; moreover, because the abundance of deuterium
varies so rapidly with $\eta$ this bound is insensitive to the
precise deuterium abundance assumed.

The difficulty in drawing sharp conclusions from the comparison between
predicted and measured primordial
abundances---e.g., stating two- or three-sigma limits---is the fact that the
dominant uncertainties, primarily in the
observations, are not Gaussian statistical errors.
Two- or three-sigma limits in the present circumstance simply have no meaning.
Instead, we choose to quote a ``sensible'' and an ``extreme''
concordance interval for the baryon-to-photon ratio.  For the
sensible interval we take $2.5\times 10^{-10}$ to $6\times 10^{-10}$,
supported from below by D+$^3$He overproduction and above by $^7$Li
overproduction.  Some have argued that $^4$He can be used to push
the upper limit down to $4\times 10^{-10}$; however, as described,
such an upper limit is exponentially sensitive to the uncertainties
in the primeval \He4 abundance, and thus not very robust.

It is interesting to compare our sensible
interval with similar concordance intervals found
by Yang et al. \cite{ytsso} in 1984, $4\times 10^{-10}\le
\eta \le 7\times 10^{-10}$, and by Walker et al. \cite{wssok}
in 1991, $2.8\times 10^{-10} \le\eta\le 4\times 10^{-10}$.  The difference
between our lower limit and that of Yang et al. is
the fact that we have allowed for slightly more astration of \He3.
The somewhat larger difference between our upper limit and
that of Walker et al. involves \Li7:  They used the lower
Spite and Spite value for the primordial \Li7 abundance and
did not allow for systematic error.  In any case, the fact
that the differences between the concordance intervals
are small is reassuring.

In setting the extreme range, we take account of
of our less than perfect understanding of the chemical evolution
of the Universe during the 10\,Gyr or so
since primordial nucleosynthesis, as well as other possible
systematic errors.  Though there is no plausible reason for
believing that $^3$He could be
astrated significantly, or that the primeval $^7$Li abundance
is significantly different from that seen in halo pop II stars,
we do not believe the wider interval of $\eta = 1.6\times
10^{-10}$ to $9\times 10^{-10}$ can be excluded with absolute
certainty.  Our extreme range derives exclusively from deuterium:
the present abundance, D/H$\,\ga 1.6\times 10^{-5}$, and
the limit to the primeval abundance, D/H$\,\la 2.5 \times 10^{-4}$.
Moreover, for such extreme values of $\eta$ all the light-element
abundances are pushed to almost untenable values.

Based upon these concordance intervals---sensible and extreme---we
can obtain bounds to the baryonic fraction of critical density,
albeit at the expense of additional dependence upon the Hubble constant,
\begin{eqnarray}
{\rm sensible:}\quad 2.5\times 10^{-10} \le \eta \le 6 \times 10^{-10}\ &
\Rightarrow & \ 0.009\le 0.009h^{-2} \le \Omega_B \le 0.02h^{-2}\le 0.14
\nonumber\\
{\rm extreme:}\quad 1.6\times 10^{-10} \le \eta \le 9\times 10^{-10}\  &
\Rightarrow & \ 0.006 \le 0.006h^{-2} \le \Omega_B \le 0.03h^{-2} \le 0.21
\nonumber
\end{eqnarray}
where the outer limits to $\Omega_B$ allow for $0.4\le h\le 1$ (see
Fig.~6).

The implications of these bounds for cosmology
are manifold and very significant.  First and
foremost, the nucleosynthesis limit is pivotal to the case for
both baryonic and nonbaryonic dark matter.  The nucleosynthesis
determination of the baryonic fraction of critical density
taken together with the observational data that indicate
that luminous matter contributes much less than 1\% of critical
density and that the total mass density is greater than 14\%
of critical density makes the case for these two most pressing
problems in cosmology.  (The Hubble-constant dependence of the
nucleosynthesis limits precludes addressing both problems by simply
choosing the right value for the Hubble constant since both the
upper and lower limits to $\Omega_B$ scale in the same way.)

Second, one can exploit the relatively well known baryon density
to estimate the total mass density by using measurements of
the ratio of total mass-to-baryonic mass in clusters of galaxies,
as determined by recent x-ray studies made using the ROSAT satellite.
Assuming that rich clusters like
Coma provide a fair sample of the ``universal mix'' of matter (if some, or all,
of the nonbaryonic dark matter is neutrinos of mass $5\eV$ to $30\eV$,
this might not be the case) and
that the hot, x-ray emitting gas is in virial equilibrium supported
against gravity only by its thermal motion, White et al. \cite{wnef}
infer a total mass-to-baryonic mass ratio of $(20\pm 5)\,h^{3/2}$,
which leads to the following estimate for the total mass density:
\begin{equation}
\Omega_0 = {M_{\rm TOT}({\rm Coma})
\over M_B({\rm Coma})}\Omega_B \simeq (0.15 - 0.5)\,h^{-1/2} .
\end{equation}
If $h$ is near the lower extreme of current measurements
this determination of $\Omega_0$ lends some support to
the theoretically attractive notion of a flat Universe (i.e., $\Omega_0 = 1$).

The ``cluster-inventory'' estimate of $\Omega_0$ is a new and
potentially very powerful method for estimating the mean density
of the Universe.  There are still important systematic sources
of error and a key assumption.  The key assumption is that
the baryons are either in stars (visible matter) or hot, x-ray
emitting gas (by a wide margin, the baryons in the hot gas outweigh
those in stars).  If there is a large amount
of baryonic matter hidden in dark stars, then $M_{\rm TOT}/M_B$ would
be smaller.  On the other hand, essentially all
systematic sources of error go in the direction of increasing
$M_{\rm TOT}/M_B$.  For example, if the hot gas is partially supported
by magnetic fields or bulk motion of the gas, then $M_{\rm TOT}$ would
be larger.  There is some evidence that $M_{\rm TOT}$ has been
underestimated:  The measurement of the mass of one cluster of galaxies based
upon weak-gravitational lensing of galaxies behind the cluster
yields a mass than is almost a factor of three
larger than that based upon x-ray studies \cite{fahlman}.
If the hot gas is clumpy, rather than smooth as is
assumed, then the gas mass would be smaller, which
also increases $M_{\rm TOT}/M_B$.  It is intriguing that a
factor of two or three increase in $M_{\rm TOT}/M_B$ would
bring the estimate of $\Omega_0$ close to unity.
In any case, further study of the x-ray and gravitational
lensing data as well as the better x-ray temperatures that the ASCA
satellite is now providing should help refine this new technique
for estimating the mean density.

While the primary concern of this paper is the baryon
density of the Universe,
big-bang nucleosynthesis also places an important constraint to
the number of light particle species present around the time
of nucleosynthesis, usually quantified as a limit to the equivalent
number of neutrino species, $N_\nu$.  This limit arises because more
species lead to additional \He4 production \cite{nulimit}.
The limit to $N_\nu$ relies upon a lower limit to $\eta$
and an upper limit to $Y_P$.  Using the D + \He3 bound to $\eta$,
$\eta \ge 2.5 \times 10^{-10}$, and our reasonable upper limit to
$Y_P\le 0.243$, it follows that  $N_\nu \le 3.4$.\footnote{Kernan
and Krauss \cite{kk} point out that by using the correlations between
the theoretical uncertainties in \He4 and D + \He3,
one can improve this limit very slightly, by about 0.1 neutrino
species, the equivalent of reducing $Y_P$ by about 0.001.}
This limit depends upon the upper limit to $Y_P$.  However, in
contrast to the upper limit to $\eta$ based upon $Y_P$,
the dependence is linear, not exponential, $N_\nu \la 3.4 + (Y_P^{\rm max}
- 0.243)/0.012$, and so the limit to $N_\nu$ is far less
sensitive to $Y_P$ than is upper bound to $\eta$.

Finally, what does the future hold?  We believe that
primordial nucleosynthesis is the best method for determining
the mean baryon density, and, in that regard, that deuterium is the
best baryometer.  Not only does the primeval deuterium
abundance vary rapidly with the baryon-to-photon ratio
(as $\eta^{-1.5}$ in the relevant range),
but prospects for measuring its primeval abundance
in high-redshift, metal-poor quasar absorption line systems look promising.
While the present situation is unsettled, with a
reported detection of $\D/\H \sim 2 \times 10^{-4}$, as well as
an upper limit, $\D /\H \le 6\times 10^{-5}$,
the situation should improve.  A handful of such
measurements could establish the primeval D abundance to
an accuracy of 10\%, which would determine the baryon density to better
than 5\% (taking account of both the observational and
theoretical uncertainties).  Because of their weak dependence
upon the baryon-to-photon ratio, as well as lingering systematic
uncertainties, $^4$He and $^7$Li are destined to play a supporting
role, albeit a very important one.  It is both ironic and
satisfying that after twenty years deuterium is still the best
baryometer.

More than forty years have passed since Gamow's introduction of the
notion of big-bang nucleosynthesis, and thirty years have passed since
the cosmic background radiation was discovered.
After more than two decades of careful comparison of theory
with observation, primordial nucleosynthesis has become
the earliest and most important test of the standard
cosmology.  Further, it leads to the best measurement
of the density of ordinary matter in the Universe and provides
a powerful laboratory for studying both the early Universe and
fundamental physics.

\section*{Acknowledgments}

We are pleased to thank Brian Fields, Keith Olive, Robert Rood,
Evan Skillman, and Gary Steigman
for helpful conversations.  This work was supported in part by the DOE
(at Chicago and Fermilab), by the NASA through grant NAGW-2381 (at Fermilab)
and grant NAGW-1321 (at Chicago), and by the NSF through grant
AST 90-22629 (at Chicago).

\newpage

\section*{Figure Captions}

\bigskip
\noindent{\bf Figure 1:}  The predictions of big-bang
nucleosynthesis.  The broken curves indicate the $2\sigma$ theoretical
uncertainties based upon our Monte-Carlo analysis.  The \He4 abundance
is given as mass fraction; the other abundances are number relative
to hydrogen.  The boxes indicate the range of baryon-to-photon
ratio consistent with the primeval light-element abundances; the
\He4 box is dotted to remind the reader that \He4 has not been
used to derive an upper limit to $\eta$ because
of the exponential dependence of such a limit to $Y_P$
(see text).  Our sensible concordance
range, $2.5\times 10^{-10} \le \eta \le 6\times 10^{-10}$,
comes from D + \He3 and \Li7 overproduction.

\medskip
\noindent{\bf Figure 2:}  The nuclear reaction network used for big-bang
nucleosynthesis; the most important reactions are numbered.
The broken boxes for mass 5 and 8 indicate that all
nuclides of this mass are very unstable.

\medskip
\noindent{\bf Figure 3:}  The \He4 mass fraction vs. nitrogen
abundance for very metal-poor, extragalactic \hii\ regions.
The solid line is the best fit to the data from the analysis of Olive
and Steigman~\cite{os}.  (In deriving their fit they
did not include some of the higher metallicity \hii\ regions).

\medskip
\noindent{\bf Figure 4:}  The $^7$Li abundance as a function
of surface temperature for very metal-poor, pop II halo stars.
The decreasing \Li7 abundance in the stars with lower surface
temperatures indicates they have burned some of their \Li7
(consistent with the fact that such stars are predicted to
have deeper convection zones).
The solid and broken lines indicate the Thorburn and Spite
plateaus respectively.

\medskip
\noindent{\bf Figure 5:}  The \Li7 abundance as a function
of iron abundance (relative to that seen in the solar system)
for stars with surface temperatures greater than 5600\,K.
The increase in \Li7 abundance seen for the stars with
higher iron abundance is indicative of additional \Li7
due to cosmic-ray processes and stellar production.
The solid and broken lines indicate the Thorburn and Spite
plateaus respectively.  For comparison, the abundances of
beryllium and boron in metal-poor, pop II halo stars are
also shown (from Refs.~\cite{beb}).  Unlike the \Li7 abundance,
the B and Be abundances increase with increasing metal abundance,
indicative of post-big-bang production.

\medskip
\noindent{\bf Figure 6:}  The fraction of critical density
contributed by baryons as a function of the Hubble constant
for the sensible concordance range of baryon-to-photon ratio (solid)
and extreme concordance range (dotted).

\end{document}